\documentstyle [12pt,epsf,rotate] {article}
\newcommand{\nwc}{\newcommand}
\nwc{\be}  {\begin{equation}}
\nwc{\ee}  {\end{equation}}
\nwc{\ba}  {\begin{array}}
\nwc{\ea}  {\end{array}}
\nwc{\bdm} {\begin{displaymath}}
\nwc{\edm} {\end{displaymath}}
\nwc{\bea} {\be\ba{rcl}}
\nwc{\eea} {\ear\ee}
\nwc{\bear} {\begin{eqnarray}}
\nwc{\ear} {\end{eqnarray}}
\nwc{\Tr} {\rm Tr}
\nwc{\LL} {\cal L}  
\begin{document}
\begin{titlepage}
\quad\\
\vspace{1.8cm}

\begin{center}
{\Large Connection between Chiral Symmetry Restoration }\\
\medskip
{\Large and Deconfinement}\\
\vspace{2cm}
C. Wetterich\footnote{e-mail: C.Wetterich@thphys.uni-heidelberg.de}\\
\bigskip
Institut  f\"ur Theoretische Physik\\
Universit\"at Heidelberg\\
Philosophenweg 16, D-69120 Heidelberg\\
\vspace{3cm}

\end{center}

\begin{abstract}
We propose a simple explanation for the connection between chiral
symmetry restoration and deconfinement in QCD at high
temperature. In the Higgs description of the QCD vacuum both
 spontaneous chiral symmetry  breaking and
effective gluon masses are generated by the condensate of a color octet
quark-antiquark pair. The transition to the high temperature
state proceeds by the melting of this condensate. Quarks and gluons
become (approximately)
massless at the same critical temperature. For instanton-dominated
effective multiquark interactions and three light
quarks with equal mass we
find a first order phase transition at a critical temperature around
170 MeV.

\end{abstract}
\end{titlepage}

\section {Octet melting}

Lattice simulations for QCD at high temperature \cite{LATB}
find empirically that chiral
symmetry restoration and deconfinement occur at the same critical
temperature
$T_c$ \cite{lattice}. Sufficiently below $T_c$ the equation of state
is rather well approximated by a gas of mesons. Gluons and quarks
or baryons play no important role. Above $T_c$ the dominant
thermodynamic
degrees of freedom are gluons and quarks. The change in the
relevant number of degrees of freedom occurs rather rapidly and may
be associated with a thermodynamic phase transition. \cite{MO}

The transition to the high-temperature state of QCD
is often called ``deconfinement transition'' since
the gluon modes with momenta $p^2\approx
(\pi T)^2$ characteristic for a thermal state behave at high temperature
close to a gas of free particles. (For experimentally
accessible temperatures above the phase transition substantial
corrections to the equation of state of a relativistic gas
remain present.) This also holds for the fermionic degrees of
freedom which appear essentially as free quarks. Having
a closer look the picture of
weakly interacting gluons is actually somewhat
oversimplified. In fact, the long-distance
behavior of the time-averaged correlation functions in high temperature
QCD corresponds to a ``confining'' three-dimensional effective
theory \cite{AP} with a strong effective coupling and a
temperature-dependent
``confinement scale'' \cite{ReuterW}, \cite{OTP}.
Furthermore, the transition between confinement and
weakly interacting ``free'' quarks is less dramatic
than it may seem at first sight. In particular, it needs
not to be associated with a true phase transition. In presence of light
quarks, the temperature-induced
changes in the long-distance behavior of the heavy
quark potential are only quantitative \cite{KLP}. String breaking
occurs for arbitrary temperature.

Nevertheless, one expects a temperature range where the
properties of strong interactions change rapidly,
reflecting ``deconfinement''. We associate here
``deconfinement'' with the transition to a thermal
equilibrium state for which (1) the bosonic contribution to the free
energy can be approximated by the one of a free gas of eight
massless spin-one bosons and, correspondingly, (2)
possible masslike terms in the effective
``gluon-propagator'' are of the order of the temperature or smaller.
We do not deal in this paper with temperature-dependent changes
in the heavy quark potential and temperature effects in the dynamics
of string breaking.

On the other hand, the chiral
properties of QCD are also expected to undergo a rapid change
at some temperature. The vacuum is characterized by a chiral
condensate of quark-antiquark pairs. This order is destroyed
at high temperature and chiral symmetry is restored. For vanishing
quark masses this implies a true phase transition. In this case
chiral symmetry restoration at high temperature
is signalled by a vanishing order parameter. At high temperature
the massless quarks contribute to the free energy as a relativistic
gas with $3N_f$ fermions whereas below a critical temperature the
fermionic degrees of freedom appear as massive baryons without
much relevance for thermodynamics.
This qualitative change in the fermionic contribution remains true
also in presence of small nonvanishing quark masses. We associate
the chiral transition at high $T$ with a rapid decrease of jump
in the chiral order parameter and the appearance of a relativistic
quark gas contribution to the free energy.

The qualitative feature that at high temperature deconfinement and
chiral
symmetry restoration come in pair is not much of a mystery: modes with
momenta $\sim T$ dominate the thermal state and the interactions between
these
modes become comparatively weak. Symmetry restoration at high $T$ is a
common
phenomenon. It is a longstanding puzzle, however, why deconfinement and
chiral
symmetry restoration occur quantitatively at the
same temperature. Analytical approaches that have been investigated
so far either concentrate on the quark degrees of freedom -- for example
in the context of Nambu-Jona-Lasinio models \cite{NJL}.
This reflects important
aspects of the chiral transition but fails to describe
the effects of confinement
and the thermodynamic equation of state of QCD. Other models deal
with pure QCD without including the important aspects connected
with the presence of light quarks. So far we are not aware of any
successful quantitative analytical description
accounting simultaneously for the
deconfinement and chiral symmetry restoration aspects of the QCD
phase transition. In view of the vaste experimental efforts \cite{EXP}
and the
important progress in numerical simulations \cite{LAT}, \cite{KLP}
even an oversimplified
analytical description of the most salient characteristics of this
transition would be very welcome.

One of the main difficulties for any analytical description are the
different
relevant degrees of freedom above and below the critical temperature.
Whereas
for high T a gas of quarks and gluons becomes a reasonable
approximation, the
low temperature physics is described by an interacting pion gas with far
less
degrees of freedom. The fermionic degrees of freedom at low T are
baryons which are
Boltzmann-supressed. Any quantitative analytical description must be
able to describe both
gluons and pions as well as quarks and baryons simultaneously. This is
mandatory at least
in the vicinity of the critical temperature which is precisely
characterized by the
transition from one set of dominant degrees of freedom to another. The
recently proposed
Higgs-picture of the QCD-vacuum \cite{CW, BerW} offers such a
possibility since all relevant
degrees of freedom are described at once. A reasonable picture of the
QCD-phase transition
may also serve as a test for these ideas.

In this paper we propose a simple explanation of the simultaneous
occurrence of the deconfinement and chiral transition and present
first quantitative estimates. Our explanation is
based on the proposed new understanding of the QCD-vacuum by
``spontaneous breaking of color'' or gluon-meson duality
\cite{CW}. In this picture of the vacuum the condensate $<\chi>$ of a
color octet quark-antiquark pair leads to spontaneous breaking of the
color symmetry. (As for the electroweak standard model there exists
an equivalent gauge invariant formulation.) The gluons acquire
a mass $\sim<\chi>$ by the Higgs mechanism and become integer charged.
Gluon-meson duality associates the massive gluons with the physical
vector mesons $\rho,K^*$ and $\omega$. This is possible since they carry
the appropriate integer electric charges as well as isospin and
strangeness.
Also the quarks carry integer charges after spontaneous color
symmetry breaking and can be associated with the baryons
$(p,n,\Lambda,\Sigma,
\Xi)$ plus a heavy singlet. They become massive due to the
spontaneous breaking of the chiral
flavor symmetry by $<\chi>$ and a similar singlet $\bar qq$-
condensate. This association between quark fields and baryons
is called quark-baryon duality. It is the analogue of color-flavor
locking
\cite{CFL} for a high baryon density.

It has been argued on phenomenological grounds
\cite{CW} that the dominant
contribution to spontaneous chiral symmetry breaking arises
from the octet condensate $<\chi>$. The same condensate therefore
explains both confinement and spontaneous chiral symmetry breaking.
In this description
the solution to the puzzle why deconfinement and chiral symmetry
restoration are connected becomes obvious. At some critical temperature
$T_c$ the value of the octet condensate $<\chi>$ drops rapidly. At this
temperature the mass of the gluons therefore drastically decreases
and the deconfinement transition happens. In the limit of three
massless quarks the expectation value $<\chi>$ exactly vanishes
at $T_c$, reflecting chiral symmetry restoration.
Typically, the color-octet and -singlet
$\bar qq$-expectation values  influence each other.
In particular, an octet condensate always induces
a singlet condensate. It seems plausible that
both the octet and singlet condensates vanish simultaneously
for $T>T_c$. Then chiral symmetry gets completely restored
for $T>T_c$, leading to a true phase transition characterized
by the order parameter of chiral symmetry breaking. The critical
temperature for the phase transition is the ``melting temperature''
for the octet condensate. The ``deconfinement temperature''
and ``chiral symmetry restoration temperature'' are
therefore identical. As a consequence of the change in the order
parameter the quantum numbers of the excitations
in the ``hadronic phase'' for $T<T_c$
differ from the ones in the ``quark-gluon phase'' for $T>T_c$.

Quark mass effects modify the details
of the transition.
For a more quantitative description of ``octet melting'' we omit
in this paper the difference between the current quark masses
and consider three light quark flavors with equal mass.
For our quantitative analysis we will need
information about the condensates in the vacuum which
are related to the form of the effective quark interactions.
We investigate
a rather wide class of effective multiquark interactions induced
by instanton effects. In the chiral limit of
three massless quarks we
find a first-order transition with critical temperature $T_c\approx
130-160$ MeV. In this limit the mass of the eight pseudoscalar
Goldstone bosons
$(\pi,K,\eta)$ vanishes at low temperature, $m_{PS}=0$. For
realistic current quark masses corresponding to a nonzero
average pseudoscalar mass
$m^2_{PS}\approx(2M_K^2+M_\pi^2)/3$,
we again get a first-order transition with a somewhat
larger $T_c\approx$ 170-180 MeV.

In this note we do not deal explicitly with
the color-singlet quark-antiquark pair which plays a subdominant
role. The relevant thermodynamic potential is then given by the
temperature-dependent effective potential for the octet condensate
\be\label{1}
U(\chi,T)=U_0(\chi)+\Delta U(\chi,T)\ee
Here $U_0$ encodes in a bosonic language for
scalar $\bar qq$-composites the information about
the multiquark interactions in the vacuum, whereas
$\Delta U$ accounts for the thermal fluctuations.
We use a scalar field $\chi_{ij,ab}\sim\bar\psi_{jb}
\psi_{ai}-\frac{1}{3}\bar\psi_{kb}
\psi_{ak}\delta_{ij}$ for the octet quark-antiquark
pair with $i,j=1...3$ color indices and $a,b=1...N_f$ flavor indices
for massless (or light) quarks. For $N_f=3$ it is sufficient to evaluate
the
effective potential for the direction of the condensate \cite{CW}
\be\label{2}
<\chi_{ij,ab}>=\frac{1}{\sqrt6}\chi(\delta_{ia}\delta_{jb}
-\frac{1}{3}\delta_{ij}\delta_{ab})\ee
In the limit of three equal quark masses this condensates preserves
a vector-like $SU(3)$-symmetry which contains the
generators of isospin and strangeness.

We emphasize that in our picture the thermodynamic quantities
depend on the number of light flavors in an important
way. For example, the vacuum condensates for two-flavor QCD ($N_f=2$)
are discussed in \cite{BerW} and differ substantially from the
three-flavor case. Needless to say that in absence of light quarks the
chiral aspects of the phase transition are completely different.
We restrict our discussion here to the realistic case
$N_f=3$. Quark mass effects are taken into account except for the
$SU(3)$-violation due to the mass differences. The
thermodynamic quantities of interest can then be extracted from
the behavior of the minima of $U(\chi,T)$. For $T=0$ the minimum
occurs for a nonvanishing octet condensate $\chi_0\not=0$. As
the temperature increases, the thermal fluctuations induce a new
local minimum at the origin $\chi=0$. At the critical temperature
$T_c$ the two minima are  degenerate. The discontinuity
characteristic for a first-order phase transition corresponds
to the jump to the absolute minimum at $\chi=0$ for $T>T_c$.

The remainder of this paper is devoted to a more detailed estimate
of $U(\chi,T)$. We will employ a mean field-type calculation
which takes into account, nevertheless, the most important
effects of the strong interactions.
We will devote some care to the proper implementation of the mean field
calculation despite
the fact that some other approximations in our approach (like the
effective vacuum potential)
remain rather crude. The reason is that we want to describe correctly
the interacting pion gas
at low temperature within a linear formulation. The linear formulation
is necessary for a
description of symmetry restoration at $T_c$, since for $T>T_c$ the
Goldstone bosons are
combined with the $\sigma$ -scalar into linear multiplets of the flavor
group. On the other
hand, the nonlinear Goldstone excitations dominate the free energy at
low $T$ and chiral
perturbation theory gives a valid description. We will see that a linear
mean field description
easily leads to incorrect results for the nonlinear excitations unless
done with sufficient
care. Furthermore, we do not want to obscure the possibility of a second
order phase
transition by using a method which is too rough. Again, unless the
bosonic degrees of
freedom are treated with care, a mean field computation can easily
produce a spurious
first order jump even in a situation of a second order transition or a
crossover. This is
well known from the study of scalar field theories or the electroweak
phase transition.

Our paper is organized as follows:
Sect. 2 addresses the form
of the vacuum potential $U_0(\chi)$, which was
estimated previously \cite{instanton} by an instanton calculation.
In sect. 3 we discuss the dependence of the masses of the most relevant
excitations on the octet condensate $\chi$. This is the basis for our
mean field-type calculation of the temperature effects $\Delta
U(\chi,T)$
in sect. 4. A first summary of quantitative results for the chiral phase
transition is given in sect. 5 in the chiral limit of massless quarks.
For the instanton-induced potential $U_0(\chi)$ proposed in sect. 2
we find a critical temperature $T_c$=154 MeV, consistent with
lattice estimates \cite{KLP}. Sect. 6 is devoted to
the high-temperature phase and the equation of state. Sufficiently
above $T_c$ one observes a weakly interacting gas of gluons and
quarks. In sect. 7 we present analytical estimates for the chiral
phase transition. In particular, for three massless quarks the critical
temperature
is related to the mass of the $\eta^\prime$-meson and its decay constant
$f_\theta$ by
\be\label{2AA}
T^4_c=0.013 M^2_{\eta^\prime}f^2_\theta R^{-\frac{1}{4}}\ee
with $R <1$ a constant of order one. \\

We turn to the value of the chiral symmetry-breaking order
parameter in the low temperature phase in sect. 8. This needs a
meaningful treatment of the Goldstone boson fluctuations
which we propose in sect. 9. In sect. 10 we verify that our
prescriptions are consistent with chiral perturbation theory for low
enough $T$. Sect. 11 discusses the effects of nonvanishing
current quark masses. We consider equal current quark masses
which correspond to an average mass of the pseudoscalar Goldstone
bosons $M_{PS}=$390 MeV. Up to $SU(3)$-violating quark mass differences
this corresponds to a realistic situation. We find $T_c=$170 MeV,
again consistent with lattice simulations \cite{KLP}. In the low
temperature phase the Higgs contribution to the mass of the vector
mesons decreases
substantially as the temperature increases. In the vicinity of $T_c$
we find a Higgs contribution to the average vector meson mass $\approx$
300 MeV and
to the average mass for the light baryons $\approx$ 600 MeV. Those have
to be supplemented
by thermal mass contributions. A decrease of the vector meson mass
\cite{BR} may be relevant
for the dilepton spectrum observed in heavy ion collisions. Finally, we
investigate in
sect. 12 a possible lattice test of our scenario by establishing the
existence of
stable $Z_3$-vortices in the low temperature phase. Sect. 13 summarizes
our conclusions.

\bigskip
\section{Vacuum-effective potential}

The effect of the thermal fluctuations has to be compared with
the vacuum-effective potential $U_0(\chi)$ for the octet. We will not
need here many details of the precise shape of $U_0$. The most
important parameter for the determination of the critical temperature
$T_c$
for the phase transition will turn out to be the difference $U_0(0)-
U_0(\chi_0)$, with $\chi_0$ the vacuum expectation value of the octet.
To be specific, we concentrate here on an instanton-induced potential
which is motivated by the observation that instanton effects are a
plausible candidate for the dynamics that lead to a non-vanishing
octet condensate \cite{instanton}. Beyond the octet $\bar qq$-condensate
realistic QCD
exhibits also a nonvanishing
color-singlet $\bar qq$-condensate $\sigma$. In order to concentrate
our discussion on the essential aspects, we assume here that $\sigma$
is ``integrated out'' by solving its effective
field equation\footnote{The effective kinetic term and
interactions of the $\chi$-field
include therefore contributions from the color singlet
field $\sigma$ as well. In a two-field language a non-zero $\sigma$
is necessarily generated for $\chi\not=0$ by terms
$\sim\sigma\chi^2$ present in the contribution from the
chiral anomaly for $N_f=3$.}
as a functional of $\chi$. We take here $\sigma\sim \chi$ (in the
relevant range) such that all $\bar qq$ condensates are accounted
for by $\chi$.

For $N_f=3$ and small values
of $\bar qq$ the instanton effects lead to a cubic term,
$U_{an}\sim\chi^3\sim(\bar qq)^3$.
For large values of
$\chi$ the nonvanishing gluon mass $\sim\chi$  leads to an
effective infrared cutoff for the instanton interactions, resulting
in $U_{an}\sim\chi^{-11}$ \cite{instanton}. A first
quantitative estimate of the instanton-induced effective potential can
be found in \cite{instanton} and we exploit its consequences
in the present work. However, in view of the uncertainties
of this estimate we will be satisfied here with a simplified version
which respects the asymptotic properties for small and large $\chi$
\bear\label{2a}
{\rm (A)}:\  U_0(\chi)=\lambda\{&-&\frac{14}{11}
R_{an}\chi_0\chi^3(1+\frac{3}{11}
(\frac{\chi}{\chi_0})^{14})^{-1}
+\nonumber\\
&&(1-R_{an})(\chi^4-2\chi_0^2\chi^2)+\chi_0^4\}\ear
The instanton contribution is the one $\sim R_{an}$
and we have added another contribution $\sim(1-R_{an})$
which arises from non-anomalous interactions not related to
instantons, as, for example, multiquark interactions mediated by gluon
exchange. Presumably the vacuum potential is dominated by
instanton effects \cite{instanton} such that $R_{an}$ is expected
close to one. In our conventions $\chi$ is real and we have
parametrized the potential such that the absolute minimum of $U_0$
occurs
at $\chi_0$, with $U_0(\chi=0)=\lambda\chi_0^4,\lambda>0$. The additive
constant is fixed such that the pressure vanishes in the vacuum. In
absence
of quark masses this implies $U_0(\chi_0)=0$.

In order to investigate how sensitively our results depend on the
precise shape of the potential, we consider for $N_f=3$ also a potential
which only incorporates the breaking of the discrete symmetry
$\chi\to-\chi$ by the anomaly, without having the correct asymptotic
behavior
\be\label{3}
{\rm (B)}:\ U_0(\chi)=
\lambda\{R_{an}(3\chi^4-4\chi_0\chi^3)+
(1-R_{an})(\chi^4-2\chi_0^2\chi^2)+\chi_0^4\}\ee
Finally, we compare our results also to a quartic potential
\be\label{3a}
{\rm (C)}:\ U_0(\chi)=
\lambda(\chi^2-\chi_0^2)^2\ee
The quartic potential
(C) is characteristic for two-flavor QCD $(N_f=2)$ (or for three flavors
with small effects of the chiral anomaly).
We emphasize that the investigation of the potentials
(B) and (C) serves mainly the purpose of an estimate
of uncertainties or errors from the unknown
details of the shape of $U_0$ whereas we consider (A) as a more
realistic potential.

As we have already mentioned, the dominant parameter for the
characteristics
of the chiral phase transition is $U_0(0)$. This measures
in our convention
the potential difference between the chiral symmetric state
at $\chi=0$ and the vacuum at $\chi=\chi_0$. It gets a contribution
from the chiral anomaly $U_{an}(0)$ and we have in our  parametrization
\be\label{U1}
U_0(0)=R^{-1}_{an}U_{an}(0)\ee
Here we denote by $U_{an}(\chi)$ the contribution of the chiral
anomaly, i.e. the part $\sim R_{an}$ in eq. (\ref{2a})
or  (\ref{3}). The size
of the chiral anomaly in the vacuum, $U_{an}(\chi_0)-U_{an}(0)$,
is given for our simple potential (\ref{2a}) by $-U_{an}(0)$. It
is directly related \cite{CW,CWmean}
to the mass of the $\eta^\prime$-meson $M_{\eta^\prime}=$
960 MeV and its decay constant $f_\theta\approx$ 150 MeV
\be\label{U2}
U_{an}(0)=\frac{f_\theta^2M^2_{\eta^\prime}}{2N_f^2}=\frac{1}{N_f^2}\cdot
10^{-2}~{\rm GeV}^4\ee
The parameter $R_{an}$ is essentially fixed for instanton-dominated
multiquark interactions $(R_{an}\approx 1)$. This sets the scale for
$U_0$ in terms of the mass and decay constant of the
$\eta^\prime$-meson.
This is the basis for an
interesting connection between the critical temperature $T_c$ on
one side and $M_{\eta^\prime}$ and $f_\theta$ on the other side which
will
become apparent later.

The vacuum octet condensate $\chi_0$ is related to the effective
gauge coupling $g$ and the vector meson mass $\bar\mu_\rho$ by $\bar
\mu_\rho=g(\bar\mu_\rho)\chi_0$. The coupling $g(\bar\mu_\rho)$ can be
estimated
\cite{CW} from the $\rho$-decay width
into two pions or charged leptons and we choose $g$(770 MeV)=6. For
$\bar
\mu_\rho$ = 770 MeV this yields
$\chi_0\approx$ 130 MeV. (In
view of the uncertainties of this estimate one could also
treat $g(\bar\mu_\rho)$
as an unknown parameter within a certain range around the estimate
above.) The coupling $\lambda$ is now determined by
\be\label{U3}
\lambda=\frac{R^{-1}_{an}f_\theta^2 M^2_{\eta^\prime}}{2N^2_f\chi^4_0}
=\frac{g^4(\bar\mu_\rho)}{2N^2_fR_{an}}\frac{f^2_\theta
M^2_{\eta^\prime}}{
\bar\mu^4_\rho}\ee
Another characteristic quantity for the vacuum potential is the
mass of the $\sigma$ resonance
given by\footnote{This holds if the contribution of the color
singlet to the kinetic term of $\chi$ is small.}
\be\label{U4}
M_\sigma^2
=\frac{3}{8}\partial^2U_0/\partial\chi^2|_{\chi_0}\ee
Phenomenology indicates $M_\sigma\approx 500$ MeV for realistic values
of the quark masses.

\section{Effective masses}

We will evaluate the temperature-dependent part of the effective
potential
$\Delta U(\chi,T)$ in a generalized mean field-type approximation.
In a strong interaction environment reasonable results can only be
obtained if the dependence of the effective particle masses on the mean
field and the momentum scale is dealt with some care.
In particular, the strong running of the couplings should not
be neglected. Furthermore, we want to be able, in principle,
to distinguish between a first or second order phase transition.
In this respect it is crucial that the Goldstone bosons are treated
correctly, since otherwise a second-order transition may be obscured
by an insufficient approximation. This is not trivial in
a mean field treatment of a linear model and we devote some
attention to this issue in sects. 9-11.

More precisely, we include the effects of the
fluctuations of the gluons, quarks,
and pseudoscalar Goldstone bosons.
Since in the low temperature phase gluons are associated
to vector mesons and quarks to baryons \cite{CW}, these
fluctuations include all light particles. Our approximation
is formally
equivalent to a gas of non-interacting massive particles.
However, the effects of the interactions appear indirectly
through the dependence of the particle masses on the mean
field and the momentum scale. In particular, the
gluons and quarks have
$\chi$-dependent masses
\be\label{4}
\mu^2_\rho(\mu)=g^2(\mu)\chi^2\quad,\quad M^2_q(\mu)=h^2_\chi(\mu)
\chi^2\ee
where we take into account the running of the gauge
and Yukawa coupling. The running of the couplings
and the choice of the renormalization
scale $\mu$ are  specified in Appendix A. From the average
baryon mass in vacuum $\bar M_q=h_\chi(\bar\mu_\rho)
\chi_0=1150$ MeV we extract\footnote{Note that actually only
$M_q(\chi=0)=0$ and $M_q(\chi=\chi_0)=\overline M_q$ are
known. The $\chi$-dependence of $M_q(\chi)$ could be more
complicated than the ansatz used in (\ref{4}).}
$h_\chi(\bar\mu_\rho)=9$.
Beyond the running couplings further effects of the interactions
of the gluons are omitted. We neglect here, in particular, the
thermal mass corrections which contribute $\sim gT$ to a chirally
invariant quark mass
and the gluom mass. (For simplicity, we also have
neglected the mass splitting between the baryon octet
and the singlet which are described by the nine
quarks.)

For the Goldstone bosons the squared mass is given
for all $T$ by the derivative of the potential
\be\label{9}
M^2_G=c_GU^\prime=c_G\frac{\partial U}{\partial(\chi^2)}
=\frac{36}{7N_f}\frac{\partial U}{\partial(\chi^2)}\ee
Also the effective meson decay constant depends on $\chi$
\be\label{5}
f^2=c_F\chi^2=\frac{7}{9}\chi^2\ee
As chiral symmetry restoration is approached for $\chi\to 0$
the effective decay constant decreases. The same holds for the chiral
condensate $<\bar\psi\psi>\sim\chi$.
We note that the general relation
\be\label{7A}
f^2M_G^2=\frac{2\chi}{N_f}\frac{\partial U}{\partial\chi}\ee
implies
\be\label{7B}
c_Gc_F=\frac{4}{N_f}\ee
independently of the individual constants\footnote{The
values used in (\ref{9}) and (\ref{5}) neglect the contribution of the
singlet
$\bar qq$-condensate.} $c_G$ and $c_F$.

\section{Thermal fluctuations}

The various contributions of the thermal fluctuations are now evaluated
as
\be\label{6}
\Delta U=\Delta_gU+\Delta_qU+\Delta_sU\ee
with
\bear\label{7}
&&\Delta_g U=24 J_B(g^2\chi^2)\ ,\quad
\Delta_qU=-12N_fJ_F(h^2_\chi\chi^2)\ ,\nonumber\\
&& \Delta_sU=
(N_f^2-1)[J_G(M^2_G)+\Delta J_G]\ear
Here we have defined the integrals
\be\label{8}
J(M^2,T)=\frac{1}{2}\left(\int^{(T)}_q\ln(q^2_T+M^2)
-\int^{(0)}_q\ln(q^2+M^2\right)\ee
with $\int^{(T)}_q=T\sum_n\int\frac{d^3\vec q}{(2\pi)^3},\ \int^{(0)}
_q=\int\frac{d^4q}{(2\pi)^4},\ q^2_T=(2\pi nT)^2+\vec q^2,\
q^2=q^2_0+\vec q^2$. For the
bosonic fluctuations
$J_B$ and $J_G$ the sum is over integer Matsubara frequencies $n$.
In contrast, for $J_F$ the Matsubara frequencies are half integer and
therefore $q^2_T=(2n+1)\pi T,\ n\in {\cal Z}$.
The integrals $J_B$ and $J_F$ reflect the difference between the one
loop expressions between
nonzero and zero temperature for a given "background field" $\chi$. They
are well
known in thermal field theory
\cite{Kap}
\bear\label{YY}
J_B(M^2,T)&=&T\int^\infty_0\frac{dqq^2}{2\pi^2}\ln(1-\exp(
-\sqrt{q^2+M^2}/T))\nonumber\\
J_F(M^2,T)&=&T\int^\infty_0\frac{dqq^2}{2\pi^2}
\ln(1+\exp(-\sqrt{q^2+M^2}/T))\ear
and can be expressed in terms of dimensionless integrals with
$m^2=M^2/T^2$
\be\label{11A}
J_{B(F)}=\frac{T^4}{2\pi^2}\int^\infty_0 d\tilde q
\tilde q^2\ln(1\mp \exp(-\sqrt{\tilde q^2+m^2}))\ee
We observe the Boltzmann suppression for large $m$, i.e.
\be\label{18AA}
\lim_{m\to\infty}J_{B(F)}=
\stackrel{-}{\scriptstyle(+)}
\frac{T^4}{2\sqrt2}\left(\frac{m}{\pi}\right)
^{3/2}e^{-m}\ee

The fluctuation integral $J_G$ for the Goldstone
bosons has to be handled with care since $M_G^2$ may be
negative for a certain range of $\chi$. The Goldstone boson mass $M^2_G$
vanishes at the temperature-dependent minimum of the
potential $\chi_0(T)$. (In our notation $\chi_0(0)\equiv\chi_0$
corresponds
to the parameter appearing in the zero-temperature effective potential
(\ref{3}).) The physics of the ``nonconvex region'' $\chi<\chi_0(T)$,
where
our approximation leads to
$\partial U/\partial(\chi^2)<0$, can be properly discussed
\cite{av} only
in the context of a coarse-grained effective action as the effective
average action \cite{Renorm, BJW}. We give here some ``ad hoc''
prescriptions
which catch the relevant physics for most (not all) situations.
For negative $U^\prime$ the fluctuations with momenta $q^2<-U^\prime$
should be
replaced by spin waves, kinks or similar ``tunneling''
configurations which fluctuate between different minima of the
potential \cite{av}. Here we simply omit the fluctuations with
$q^2<-U^\prime$.
As a consequence, $J_G$ equals $J_B$ for $U^\prime\geq0$, whereas
the lower integration boundary for $\tilde q$ in eq. (\ref{11A}) is
replaced by $|m|$ if $m^2$ is negative. With
\be\label{11A1}
J_G(M^2,T)=\left\{\begin{array}{lll}
J_B(M^2,T)&{\rm for}& M^2\geq0\\
\frac{T^4}{2\pi^2}\int^\infty_0dyy\sqrt{y^2+|m^2|}\ln(1-e^{-y})&
{\rm for}& M^2<0\end{array}\right.\ee
we observe that $J_G$ is continuous in $M^2$. For $M^2<0$ the
contribution
of the Goldstone fluctuations is enhanced as compared to
a gas of massless free particles. Finally, the correction $\Delta J_G$
in eq. (\ref{7}) is related to the appropriate treatment of the
running couplings and specified in
Appendix A.

Our treatment of the composite scalar fluctuations leads to
some shortcomings. The perhaps most apparent one concerns the behavior
of $\Delta U$ for very high $T$. In this region the gluons have only
two helicities and the factor 24 multiplying $J_B$ should be replaced
by 16. In fact, a more correct formulation would treat the $2(N_c^2-1)$
transversal gluon degrees of freedom separately from the longitudinal
degrees of freedom. The latter ones are associated to the scalar sector
by the Higgs mechanism. In contrast to the fundamental scalar in
the electroweak sector of the standard model the octet $\chi$ is a
composite and the scalars should not be counted as independent degrees
of freedom at high $T$. In a more complete treatment this
is realized by large $T$-dependent
mass terms for all scalars which suppress their contribution effectively
for large $T$.
The neglection of this effect in our rough treatment leads to an
overestimate of the bosonic contribution\footnote{The suppression
of the Goldstone boson contribution $J_G$ at large $T$ is included
properly in our approximation.} at large $T$ by a factor 3/2.

On the other hand, in the low temperature region we have neglected
bosonic contributions from the $\sigma$-resonance or the scalar
octet ($a^\pm$ etc.) which have a similar mass as the vector mesons.
This results in an underestimate of the bosonic contribution which
is particularly serious near a second-order or weak first-order
phase transition. At a second-order phase transition all scalars
contained in an appropriate linear representation become massless.
In the chiral limit the minimal set are 144 real scalars for
$N_f=3$ and 32 real scalars  for $N_f=2$. In contrast, our treatment
accounts only for $(N_c^2-1)+(N^2_f-1)$ massless scalars at the
phase transition.

Another shortcoming is the neglection of temperature effects in
the instan\-ton-induced interactions. They become important for $\pi T
\approx \mu_\rho$ and can be neglected for low $T$. It is conceivable
that these effects influence substantially the details of the
phase transition. Despite of all that we remind that the gross
features of the equation of state and the value of the critical
temperature are relatively robust quantities. We will see below that an
error of 20 \% in the number of effective degrees of freedom in the
quark-gluon phase at $T_c$ influences the value of $T_c$ only
by 5 \%. A similar statement holds for a\\
20 \% error in the estimate
of $U_0(0)-U_0(\chi_0)$.
For a first rough picture of the implications of the color octet
condensate on the chiral phase transition our approximation of the
thermal fluctuations (\ref{6}) seems therefore appropriate.

\section{High temperature
phase transition  \protect\\
for vanishing quark masses}

\medskip
For a given form of the vacuum potential $U_0$ the temperature
dependence of the effective potential and the temperature-dependent
masses of the pseudo-particles can now be computed. We concentrate
here on a potential that is dominated by instanton effects
reflecting the chiral anomaly and take $R_{an}=0.9$
in eq. (\ref{2a}). In table 1
we present for the chiral limit(vanishing current-quark masses)
the values of the vector-meson mass $\bar\mu_\rho$ (input), the
decay constant $f$ and the $\sigma$-mass $M_\sigma$ for
the vacuum. The results for the instanton-induced
potential are denoted by (A). For our parameters realistic
values of $\bar\mu_\rho, f$ and $M_\sigma$ are obtained
for a realistic average current quark mass (see sect. 11, table 2).
They differ from the chiral limit shown in table 1.
In order to investigate the influence
of the uncertainties in the
vacuum potential on the thermal properties,
we also compare with corresponding results for
the potentials (B) (C) in eqs. (\ref{3}), (\ref{3a}).
(For the potentials (B), (C) we
have not  optimized
the input parameters $\bar\mu_\rho,
g(\bar\mu_\rho),h_\chi(\bar\mu_\rho)$
with respect to the observed particle masses for nonzero current quark
masses.)

As the temperature increases, we
find a first-order phase transition. The critical temperature
$T_c$ is indicated in table 1, as well as the Higgs contribution to the
mass of the vector
mesons and baryons slightly below the critical temperature
$(\bar\mu_\rho^{(SSB)}(T_c)$ and $M_q^{(SSB)}(T_c)$).  We observe that
the
decrease of the effective masses $\bar\mu_\rho(T)$ and $M_q(T)$ for
increasing temperature is largely determined by the temperature
dependence of the effective gauge and Yukawa couplings and only to a
smaller
extent by the decrease of the expectation value  $\chi_0(T)$. We observe
that the
true screening masses get temperature corrections which increase $\sim
T$ and
become important near $T_c$.
Also the maximum of the vector meson spectral function, which is
relevant for dilepton
production, differs from the screening mass. As $T$
increases beyond $T_c$, the absolute minimum of $U(\chi,T)$ jumps
to $\chi=0$ and the gluons and quarks (or vector-mesons and baryons)
become massless in our approximation. The equation of state approaches
rapidly the one
for a free gas of gluons and quarks (see sect. 6 for details).

Since the potentials (B)(C) are qualitatively quite
different from (A), we believe that the range $T_c\approx$ 130-160 MeV
roughly covers the uncertainty from the choice of the vacuum potential
$U_0$. We note that the critical temperature for the
instanton-induced potential
(A) comes close to the value $T_c=154\pm 8$ MeV
suggested by a recent lattice simulation \cite{KLP}. If we include only
the transversal gluon degrees of freedom for $\chi=0$ (see sect. 4) the
value
of $T_c$ becomes somewhat larger by about 4 \%.

\begin{center}
\noindent {\bf Table 1:} First-order phase transition in
the chiral limit $(m_{PS}=0)$.\\

\medskip
\begin{tabular}{|l|ccc|c|cc|c|}
\hline
\multicolumn{4}{|c|}{\rm vacuum}
&\multicolumn{4}{c|}{\rm critical temperature}
\\
\hline
$U_0$&$\bar\mu_\rho$&$f$&$M_\sigma$&$T_c$&$\bar\mu_\rho^{(SSB)}(T_c)$&
$M_q^{(SSB)}(T_c)$&$\tau(T_c)$\\
\hline
A&700&68&1570&154&290&580&0.37\\
B&770&113&580&134&440&700&0.66\\
C&770&113&480&131&450&720&0.71\\
\hline
\end{tabular}
\end{center}

\section{High temperature phase}

The results of a numerical evaluation
of $U(\chi,T)$ can be understood qualitatively by simple
analytical considerations. For
large temperature $T>M$ the integrals $J$ are dominated
by their values at $M=0$
\be\label{T1}
J_B(0,T)=J_G(0,T)=-\frac{\pi^2}{90} T^4\ ,\quad J_F(0,T)=
\frac{7\pi^2}{720}T^4\ee
If the minimum of $U$ for large $T$ occurs at $\chi=0$,
the gauge bosons and quarks are effectively massless. The pressure
is then given by
\be\label{T2}
p=-U(\chi=0)=\frac{16\pi^2}{90}T^4+
\frac{12N_f\pi^2}{90}\left(\frac{7}{8}\right)T^4-U_0(0)\ee
We recognize the contribution of a free gas of $2(N_c^2-1)=16$ bosonic
and $4N_cN_f=12N_f$ fermionic massless degrees of freedom. The scalar
part $\Delta_sU$ is subleading and will be neglected (see sect. 4).
The negative contribution $-U_0(0)=-\lambda\chi_0^4$ accounts
in our approximation for interaction effects related to the octet
condensation. It becomes irrelevant only for large $T$ whereas
for temperatures near $T_c$ its effects cannot be neglected.
For large $T$ the pressure approaches
rapidly the Stefan-Boltzmann limit
\be\label{30AX}
p_{SB}=\left(\frac{8}{45}+\frac{7}{60}N_f\right)\pi^2T^4\ee

The energy density
\be\label{T3}
\epsilon=U-T\frac{\partial U}{\partial T}=U_0(0)+\frac{8}{15}
(1+\frac{21N_f}{32})\pi^2 T^4\ee
deviates from the one for a relativistic gluon-quark gas
due to $U_0$. One obtains the equation of state
\be\label{T4}
\epsilon-3p=4U_0(0)\ ,\quad \frac{\epsilon-3p}{\epsilon+p}=
\frac{45}{(8+\frac{21}{4}N_f)\pi^2}\frac{U_0(0)}{T^4}\ee
This yields for $N_f=3$
\be\label{EQ1}
\tau=\frac{\epsilon-3p}{\epsilon+p}\approx 0.192\frac{U_0(0)}{T^4}
=\tau(T_c)\frac{T_c^4}{T^4}\ ,
\quad \frac{p}{\epsilon}=\frac{1-\tau}{3+\tau}\ee
We have indicated the value of $\tau(T_c)$
obtained from the numerical evaluation of eq. (\ref{6})
in the chiral limit in
table 1 (and for realistic quark masses in table 2 in sect. 11). The
small value of $\tau$ for the
instanton-induced potential (A) implies that
the equation of state for a relativistic plasma is
particularly rapidly
approached as the temperature increases beyond $T_c$.
The fast turnover to a relativistic plasma
is consistent with findings in lattice gauge theories \cite{lattice}.

In a more accurate treatment the free energy of the quark gluon gas and
the
Stefan-Boltzmann law receive additional perturbative corrections also
for large $T$.
The dominant corrections can be incorporated in our picture by adding in
eq. (11)
the temperature dependent mass terms \cite{QUASIP}. We emphasize,
however, that the
contribution $-U_0(0)$ in eq. (24) remains as an important
non-perturbative effect.

\section{Chiral phase transition}

For a first-order transition the critical temperature
$T_c$ can be related to $U_0(0)$. Let us denote the value of
the potential in the hadronic phase at $T_c$ by
$U_{SSB}=U(\chi_0(T_c))$. Equating this with eq.
(\ref{T2}) yields
\be\label{P1}
T^4_c=\frac{45}{8\pi^2}\left(1+\frac{21N_f}{32}\right)^{-1}
(U_0(0)-U_{SSB})\ee
where we have neglected $\Delta_sU$ since
the Goldstone boson fluctuations are not relevant
at $\chi=0$ for a sufficiently strong first-order transition.
As long as $T_c$ remains much smaller than
the mass of the gauge bosons $\bar\mu_\rho^{(SSB)}$
and fermions $M_q^{(SSB)}$ in the phase with broken
chiral symmetry, their contribution to
$U_{SSB}$ remains
strongly Boltzmann-suppressed (cf. table 1).
For a strong first-order phase transition
the quantity $U_{SSB}$ is therefore dominated by the fluctuations
of the $N_f^2-1$ Goldstone bosons. If $\chi_0(T)$ is close to
$\chi_0=\chi_0(0)$ we can also neglect the difference $U_0(\chi_0
(T))-U_0(\chi_0)$ and $U_{SSB}$ is given by the contribution
of a free gas of Goldstone bosons
\be\label{P2}
U_{SSB}=-\frac{(N_f^2-1)\pi^2}{90}T_c^4\ee
As a consequence, the critical temperature is given by $U_0(0)$ quite
independently of the details of the shape of $U_0(\chi)$ or the
thermal fluctuations
\be\label{P3}
T^4_c=\frac{45}{8\pi^2}(1+\frac{21N_f}{32}-
\frac{N_f^2-1}{16})^{-1}U_0(0)=0.231\quad U_0(0)\ee
(For the quantitative estimate we use $N_f=3$.)
For the approximation (\ref{EQ1}) this yields an equation of state
\be\label{EQ2}
\frac{\epsilon-3p}{\epsilon+p}\approx \tau(T_c)\frac{T_c^4}{T^4}
\approx0.83\frac{T^4_c}{T^4}\ee
and we see again that a free relativistic gas of gluons and quarks
is reached rapidly as $T$ increases beyond $T_c$. The pressure
is continuous at $T_c$ and directly given by
$p(T_c)=-U_{SSB}$. Comparing eq. (\ref{P2}) with the Stefan-Boltzmann
value  (\ref{30AX}) yields
$p(T_c)/p_{SB}=\frac{16}{95}\left(\frac{3}{37}\right)$ for $N_f=3(2)$.

Eq. (\ref{P3}) relates the critical temperature to the properties
of the $\eta^\prime$-meson and the fractional contribution
$R_{an}$ of the chiral anomaly to $U_0(0)$ (cf. eq. (\ref{U2}))
\be\label{P4}
T_c=R^{-1/4}_{an}N_f^{-1/2}\cdot 221\ {\rm MeV}\ee
For $R_{an}=(0.3,0.5,0.8,1)$ and $N_f=3$ this yields $T_c=
(173,152,135,128)$ MeV. Fluctuations in the SSB-phase beyond the pions
(30)
further lower the value of $U_{SSB}$. This replaces in eq. (33) $R_{an}$
by $\bar{R}$, with
$\bar{R} <R_{an}$, leading to an increase of $T_c$. For three-flavor QCD
our numerical
results are consistent with this qualitative
picture of a strong first-order phase transition.

The quantitative
estimates (\ref{P3})-(\ref{P4}) are only valid
if the minimum at $\chi_0(T_c)$ is
still near $\chi_0$. This depends on the
influence of the Goldstone bosons
and, in particular, on the ratio between $T_c$ and the meson decay
constant $f$ (see sect. 11). For a weak first-order transition or a
second-order transition (for $N_f=2$) the potential difference $U_0
(\chi_0(T_c))$ has to be included in $U_{SSB}$ (\ref{P2}). This
contribution lowers the critical temperature. On the other hand, the
gauge boson and quark fluctuations may not be negligible anymore
for low enough $\chi_0(T_c)$. This effect enhances the critical
temperature. The difference between the rough estimates (\ref{P4})
and (\ref{EQ2}) and the numerical conputation (table 1) is
related to these effects\footnote{The numerical computation
also uses 24 gauge boson degrees of freedom (eq. (\ref{7}))
instead of 16 for the analytical discussion in sects. 7 and 8. This
lowers
$T_c$ by 4 \%.}. It is particularly pronounced for the
potential (A) for which the gauge boson mass in the SSB-phase
is already considerably smaller at $T_c$ than at $T=0$.

In the two-flavor case the chiral phase transition in QCD ressembles
in many respects the electroweak phase transition. The
vacuum potential is analytic in $\chi^2$ at $\chi^2=0$ such
that a first-order transition can only be induced by the fluctuations
of the gauge bosons. In this analogy the octet replaces the
Higgs-scalar and $M_\sigma$ plays the role of the mass of the Higgs
scalar $M_H$. Also the gluons replace the weak gauge bosons
and the quarks play the role of the top quark in the electroweak
theory. The electroweak phase diagram shows a first-order phase
transition for small $M_H$
and one may expect the same for QCD for small $M_\sigma$.
In the electroweak theory this transition ends for some critical
$M_{H,c}$ in a second-order transition, whereas for
$M_H>M_{H,c}$ the transition is replaced by
a continuous crossover \cite{ReuterW}, \cite{Sintra},\cite{crossover}.
In the case of QCD
with vanishing quark masses one has, however, an additional
important ingredient, namely the existence of an order parameter
and Goldstone bosons. In fact, in
the absence of current quark masses an order parameter
$\chi_0(T)\not=0$
breaks spontaneously a global symmetry. In
consequence, there must be a phase transition as $\chi_0(T)$ reaches
zero when the critical temperature $T_c$ is reached from below.
This holds for arbitrarily large $M_\sigma$. In two-flavor QCD
with vanishing quark masses a second-order phase transition replaces
the crossover of the electroweak theory. For nonzero quark mass the
analogy is even closer, with a change from a first-order
transition to crossover in dependence on $M_\sigma$. In contrast to the
electroweak theory, however, $M_\sigma$ is not a free parameter but
can, in principle, be computed in QCD. An important ingredient
for the determination of the characteristics of the phase transition
is the size of the gauge coupling which is large in QCD. The realization
of a second-order transition/crossover seems therefore quite
plausible in two-flavor QCD.

\section{Spontaneous chiral symmetry breaking for $T<T_c$}

For an understanding of the temperature effects in the hadronic
phase for $T<T_c$ we need the value of
$\chi_0(T)$.
The octet expectation value in thermal equilibrium obeys the field
equation
\be\label{FE1}
\frac{\partial U}{\partial\chi}=2\chi\frac{\partial U}
{\partial(\chi^2)}=2\chi U^\prime=j_\chi\ee
where $j_\chi=0$ in absence of quark masses. For a determination of the
temperature dependence of the  order parameter $\chi_0(T)$ it is
useful to investigate the field equation analytically. For
the gluon contribution one finds from (\ref{7})
\be\label{10}
\Delta_gU^\prime\equiv\frac{\partial}{\partial(\chi^2)}\Delta_gU=12
\hat g^2(\mu_\rho(T))I^{(B)}_1(\mu^2_\rho(T),T)\ee
Here
\be\label{28a}
\hat g^2=\frac{\partial\mu^2_\rho(T)}{\partial(\chi^2)}
=g^2(\mu_T)\left\{1-\frac{\mu_\rho^2}{\mu_\rho^2+(\pi T)^2}
\frac{(\beta_g/g)(\mu_T)}{(\beta_g/g)(\mu_\rho)}(1-
\frac{g}{\beta_g}(\mu_\rho))^{-1}\right\}\ee
reflects the running of the gauge coupling
in dependence
on the temperature-dependent gluon mass
\be\label{28b}
\mu^2_\rho(T)=g^2(\mu^2_T)\chi^2\ee
The integral
\bear\label{11}
I_1^{(B)}(M^2,T)&=&2\frac{\partial}{\partial M^2}J_B(M^2,T)\nonumber\\
&=&T\sum_n\int\frac{d^3\vec q}{(2\pi)^3}\frac{1}
{\vec q^2+(2\pi nT)^2+M^2}-\int\frac{d^4q}{(2\pi)^4}\frac{1}{q^2+M^2}
\nonumber\\
&=&\frac{T^2}{2\pi^2}\int^\infty_0dy\frac{\sqrt{y^2+2my}}{e^{y+m}-1}
\ear
contains for large $T$ terms
quadratic, linear and logarithmic
in $T$ \cite{Dolan}
\be\label{11b}
I_1^{(B)}\approx\frac{T^2}{12}-\frac{TM}{4\pi}+\frac{M^2}{16\pi^2}\ln\frac
{(\tilde cT)^2+M^2}{M^2}\ee
More precisely, the  expression (\ref{11b}) applies for
$M^2\ll \pi^2 T^2$  and
we will drop the logarithm in the following. We observe an infrared
divergence in $\partial I_1^{(B)}/\partial M^2$
for $M\to 0$.

The fermionic contribution reads
\be\label{12}
\frac{\partial}{\partial(\chi^2)}\Delta_qU=-6N_f
\hat h^2I_1^{(F)}(
h_\chi^2\chi^2,T)\ee
with
\bear\label{30a}
\hat h^2&=&\frac{\partial}{\partial (\chi^2)}(h^2_\chi(\mu_T)\chi^2)\\
&=&
h^2_\chi(\mu_T)\left\{1-\frac{\mu_\rho^2}{\mu_\rho^2+(\pi
T)^2}\frac{(\beta_h/h_\chi)(\mu_T)}
{(\beta_g/g)(\mu_\rho)}\left(1-\frac{g}{\beta_g}(\mu_\rho)\right)^{-1}
\right\}\nonumber\ear
The corresponding integral
\bear\label{12a}
I_1^{(F)}(M^2,T)&=&2\frac{\partial}{\partial M^2}J_F
(M^2,T)\nonumber\\
&=&-\frac{T^2}{2\pi^2}\int^\infty_0dy
\frac{\sqrt{y^2+2my}}{e^{y+m}+1}=\frac{1}{24}T^2+...\ear
involves half-integer Matsubara
frequencies and does not show an infrared divergence.

The Goldstone-boson integral $I_1^{(G)}=2\partial J_G/\partial M^2$
equals
$I_1^B$ for $M^2\geq0$.
For negative $M^2$ it becomes
\be\label{14A}
I_1^{(G)}=-\frac{T^2}{2\pi^2}\int^\infty_0
dyy(y^2+|m^2|)^{-1/2}\ln(1-e^{-y})
=\frac{T^2}{12}+\Delta I_1^{(G)}\ee
For small values of $|M|/T$ one finds
\be\label{14B}
\Delta I^{(G)}_1=\frac{T|M|}{2\pi^2}\left(\ln\frac{2|M|}{T}-1\right)+...
\ee
and we see that $I_1^{(G)}$ is continuous at $M^2=0$. On the other hand,
for large $|M|/T$ the result is
\be\label{48AA}
I_1^{(G)}=\frac{\hat c}{4\pi^2}\frac{T^3}{|M|}+...\ ,
\quad \hat c=\int^\infty_0dy\frac{y^2}{e^y-1}\approx 2.4041\ee

For sufficiently large $T$ a positive temperature-dependent
mass-like term
dominates the potential derivative at the origin
\be\label{48AB}
\frac{\partial\Delta U}{\partial(\chi^2)}_{|\chi=0}=\left(
\hat g^2+\frac{N_f}{4} \hat h^2_\chi
\right)T^2\ee
This overwhelms any classical contribution and there is therefore
no nontrivial minimum of $U(\chi,T)$ for $\chi\not=0$.
Chiral symmetry is restored $(\chi_0(T)=0)$ and the
gluons do no longer acquire a mass from the octet condensate.
On the other hand, for $T<T_c$ the absolute minimum of the
potential occurs for $\chi_0(T)\not=0$, corresponding to a
nontrivial solution of eq. (\ref{FE1}).

\section{Goldstone-boson fluctuations}

For a determination of $\chi_0(T)$ in the low temperature
phase with chiral symmetry breaking the
Goldstone boson fluctuations are important. They dominate the
low-temperature behavior and play an important role at a possible
second-order phase transition. The correct treatment of the
Goldstone bosons in a mean field-type approach is rather subtle
due to complications in the infrared physics for massless
bosons. (This also applies to gauge bosons.) We show in the next
section that our renormalization-group motivated prescriptions
lead to the correct behavior at low $T$ and we demonstrate
in appendix B that they also can account\footnote{A naive
mean field treatment typically produces a first-order
transition when massless bosons are present. This is due to
the nonanalytic ``cubic term'' in the expansion of $J_B$ (eq.
(\ref{YY}) in powers of $M$.} for the possibility
of a second-order phase transition. The latter is
particularly relevant for the case of two light quark flavors or for
realistic QCD where the physical strange quark mass seems to correspond
to a very weak
first order transition or a crossover, in the vicinity of a second order
transition
for an nearby critical strange quark mass.

The Goldstone boson mass
involves the derivative of the
temperature-dependent effective potential $M_G^2=c_GU^\prime$, which
obeys
for all $T$
\bear\label{16C}
U^\prime&=&B(\chi)+\frac{N_f^2-1}{2}c_GU^{\prime \prime}\Delta
I_1^{(G)}(c_GU^\prime,T)+...\\
B(\chi)&=&U^\prime_0+\frac{N_f^2-1}{24}c_GT^2U_0^{\prime \prime}
+12\hat g^2I_1^{(B)}(g^2\chi^2,T)-6N_f\hat h^2
I_1^{(F)}(h^2_\chi\chi^2,T)\nonumber\ear
Here the dots stand for corrections arising from
$\Delta J_G$ (\ref{20A3}) which vanish for $U^\prime=0$ and are
negligible in the vicinity of an extremum.
>From $\Delta I_1^{(G)}(0,T)=0$ one infers a simple equation
for the extrema of $U$ away from the origin. The condition
$U^\prime=0$ is realized for $B=0$ or
\be\label{16E}
U_0^\prime+\frac{N_f^2-1}{24}c_GU_0^{\prime \prime} T^2=6N_f\hat h^2
I_1^{(F)}(h_\chi^2(\mu_T)\chi^2_0,T)-12\hat g^2I_1^{(B)}(g^2
(\mu_T)\chi_0^2,T)\ee
where we recall the definitions
\be\label{16F}
U_0^\prime=\frac{1}{2\chi}\frac{\partial U_0}{\partial\chi}\quad, \quad
U_0^{^{\prime \prime}}
=\frac{1}{4\chi^2}\left(\frac{\partial^2 U_0}{\partial\chi^2}
-\frac{1}{\chi}\frac{\partial U_0}{\partial \chi}\right)\ee
For low temperature the r.h.s. of eq. (\ref{16E}) is
Boltzmann-suppressed
and can be neglected.

The differential equation (\ref{16C}) for $U^\prime(\chi)$ can be turned
into
an algebraic ``Schwinger-Dyson equation'' by
replacing $U^{^{\prime \prime}}$ by $U_0^{^{\prime \prime}}$. This is
reasonable for
$U_0^{\prime \prime}\geq0$, whereas for negative $U_0^{\prime \prime}$
we omit the contribution $\sim U_0^{\prime \prime}$ in $B$
and $\Delta I_1^{(G)}$ in eq. (\ref{16C}) such that $U^\prime=B$. These
prescriptions\footnote{In regions of the potential where the
gauge boson and fermion fluctuations are important a
better approximation would replace $U_0^{\prime \prime}\to U_0^{\prime
\prime}+\Delta_gU^{\prime \prime}+
\Delta_qU^{\prime \prime}$, similar to eq. (\ref{23AA}).} determine
$U^\prime$ for given values of $\chi$ and $T$. Not too far from the
minimum one may further approximate (for $U_0^{\prime \prime}\geq0$)
\be\label{16D}
U^\prime=B-\frac{N_f^2-1}{8\pi}
c_G^{3/2}U_0^{\prime \prime} T\left(\sqrt{U^\prime\Theta(U^\prime)}
-\frac{2}{\pi}\sqrt{-U^\prime\Theta(-U^\prime)}
(\ln(2|U^\prime|/T)-1)\right)\ee
A reasonable approximate formula for $U^\prime$ used for our numerical
computation is given by
\bear\label{16G}
U^\prime&=&\frac{1}{4}sign(B)(\sqrt{E^2+4|B|}-E)^2,\nonumber\\
E&=&\frac{N_f^2-1}{8\pi}c_G^{3/2}TU_0^{\prime \prime}\Theta(U_0^{\prime
\prime})\ear
This implies that for small $B$ in the vicinity of the minimum where
$B\sim\chi-\chi_0$, one has\footnote{This holds for
$E>0$. Note that for $U^\prime\sim
\chi-\chi_0$ the nonanalyticity in $\Delta I^{(G)}_1$ would
destabilize the minimum.} $U^\prime\sim(\chi-\chi_0)^2$.

\section{Low temperature pion gas and chiral perturbation theory}

The thermodynamics for low temperature should be described by a gas of
interacting
pions. Any analytical computation which pretends to describe the
transition from
a pion gas at low $T$ to a quark gluon plasma at high $T$ should
reproduce the low temperature
limit correctly. Indeed, in our formulation only the massless or very
light particles
contribute for low $T$. In particular, around the minimum of $U$ near
$\chi_0$
both the baryons and the vector mesons (or, equivalently, the quarks
and the gluons) are heavy and exponentially Boltzmann-suppressed. Only
the
pions are light and one expects that they completely dominate the
temperature effects at low $T$ for $\chi$ near $\chi_0$.

The effects of a thermal interacting
pion gas are described by chiral perturbation theory
\cite{CHPT} which predicts for the chiral condensate\footnote{Here
we have neglected the $T$-dependence of the wave function
renormalization which relates $\chi$ and $<\bar\psi\psi>$.}
\be\label{LT1}
\frac{\chi_0(T)}{\chi_0(0)}=\frac{<\bar\psi\psi>
(T)}{<\bar\psi\psi>(0)}=1-\frac{N_f^2-1}{N_f}\frac{T^2}{12 f^2}-
\frac{N_f^2-1}{2N_f^2}\left(\frac{T^2}{12f^2}\right)^2+...\ee
Within the nonlinear setting of chiral perturbation theory
one can understand the temperature effects of the interacting
pions on the chiral condensate in a staightforward way. In
particular, it is obvious that the lowest two orders can only depend
on the ratio $T/f$ \cite{CHPT}. In order to describe
the phase transition and the high temperature phase we are
bound, however, to use a linear description for the scalar fields.
The correct reproduction of lowest order chiral perturbation
theory is not trivial in a linear setting and can be used
as a test for the correctness of the particular formulation of
the meanfield approximation.

In
our picture we can parametrize the effective potential in
the vicinity of the minimum $\chi_0$ by $(U=U_0+\Delta U)$
\bear\label{LT2}
U_0&=&\frac{4}{3}M^2_\sigma(\chi-\chi_0)^2\nonumber\\
\Delta U&=&(N_f^2-1)
\left(-\frac{\pi^2}{90}T^4+\frac{c_G}{24}T^2\partial U_0/\partial
(\chi^2)\right)\ear
Up to the $\chi$-independent contribution $\sim T^4$ this yields
\be\label{LT3}
U=\frac{4}{3}M^2_\sigma\{(\chi-\chi_0)^2+\frac{c_G(N_f^2-1)}{24}\frac
{T^2}{\chi}(\chi-\chi_0)\}\ee
and one infers that the $T$-dependent minimum is independent
of $M_\sigma$
\be\label{LT4}
\frac{\chi_0(T)}{\chi_0(0)}=1-\frac{c_Gc_F}{4}(N_f^2-1)
\frac{T^2}{12f^2}\frac{\chi_0^2(0)}{\chi^2_0(T)}\ee
With $c_Gc_F=4/N_f$ (cf. eq. (\ref{7B}))
we can verify that the lowest order of an expansion
in $(\chi_0(T)-\chi_0(0))/\chi_0(0)$ corresponds indeed
to chiral perturbation theory. This shows the consistency
of our treatment of the scalar fluctuations as discussed
in Appendix A. A lack of care in the
treatment of the scalar fluctuations easily leads to inconsistency
with chiral perturbation theory even in lowest order. In fact, it is
crucial that the term $\sim T^2$ in $\Delta U$ (eq. (\ref{LT2})
involves the derivative of $U_0$ without additional temperature
corrections to the potential.

\section{Nonzero quark masses}

We finally discuss the effect of the quark masses in the
limit where they are all equal. A quark mass term adds to the potential
a linear\footnote{We omit here nonlinear quark mass corrections
from the chiral anomaly. We also observe that the
linear term actually appears for the $\bar qq$-color singlet $\sigma$
and is transmuted to $\chi$ only by integrating
out $\sigma$. This results again in a nonlinearity of $U_m$. We consider
$U_m$ as an approximation for $T\leq T_c$. There is actually
only a singlet and no octet condensate in the high
temperature phase for nonvanishing quark masses.} piece
\be\label{48AC}
U_j=U_0+U_m\quad , \quad U_m=-j_\chi(\chi-\hat\chi_0)\ee
Therefore the location $\bar\chi_0$ of the minimum of $U+U_m$ obeys
for arbitrary $T$
\be\label{48AD}
\frac{\partial U}{\partial\chi}(\bar\chi_0)=j_\chi\ee
Here $j_\chi$ is proportional to the
current quark mass and can be related to the mass of the octet of light
pseudo-Goldstone bosons $(\pi,K,\eta)$ by
\be\label{48AE}
m^2_{PS}(T)=c_G\frac{\partial U}{\partial(\chi^2)}(\bar\chi_0(T))=
\frac{c_G}{2\bar\chi_0(T)}j_\chi\ee
In particular, $\bar\chi_0=\bar\chi_0(0)$is determined by the vacuum
mass
$m_{PS}=m_{PS}(0)$
\be\label{48AF}
U_0^\prime(\bar\chi_0)=m^2_{PS}/c_G\ee
Vanishing pressure in the vacuum (at $T=0$)  requires $U_0(\bar\chi_0)
+U_m(\bar\chi_0)=0$ or
\be\label{48AG}
\hat\chi_0=\bar\chi_0-U_0(\bar\chi_0)/j_\chi\ee
For a small quark mass (small $j_\chi$) both $\bar\chi_0$ and
$\hat\chi_0$
are given approximately by $\chi_0$.

Near the chiral limit we can understand various
quark mass  effects on the chiral phase transition analytically:

(1) The vacuum expectation value $\chi_0$ increases and $U_0(0)
+U_m(0)$
acquires an additional positive contribution $j_\chi\hat\chi_0$. Both
effects enhance $T_c$.

(2) In the high-temperature phase the minimum occurs at
$\chi_s(T)\not=0$.
For a first-order transition this reduces the
difference between $U_j$ in the low and high
temperature phases and therefore diminishes somewhat the
increase of $T_c$. Also the  quarks get a nonzero
mass in the high-temperature
phase,  $h_\chi\chi_s(T)$. This decreases
the pressure of the fluctuations at given $T$ and therefore
enhances the value of $T_c$ needed in order to compensate
the difference in $U_j$. For small $j_\chi$ this effect is only
quadratic in $j_\chi$ and the linear effect in
$U_j(\chi_s(T))-U_j(\chi_0(T))$ dominates.

(3) As $T_c$ and $\chi_s(T_c)$ increase with increasing $j_\chi$, the
difference between the two local minima characteristic for the
first-order
transition becomes less and less pronounced. Finally, the transition
line ends for a critical $j_{\chi,c}$ at a second-order transition,
with crossover for $j_\chi>j_{\chi,c}$. The situation is similar
to the liquid-gas transition.

(4) In case of a second-order transition for $j_\chi=0$ the chiral
transition turns to a crossover for all nonzero $j_\chi$.

In summary, nonvanishing quark masses enhance the critical temperature
and make transitions less pronounced. The case of realistic QCD with
$m_s\not=m_{u,d}$ is somewhat more complicated since the expectation
values in the high temperature phase differ in the strange and
nonstrange directions. In case of a strong first-order transition
this modification is, however, only of little quantitative relevance.
On the other hand, one may envisage a situation where for $m_s\not=0,
m_u=m_d=0$ the critical behavior ressembles two-flavor QCD.

Our numerical evaluation of $U(\chi,T)$ is easily exploited
for nonvanishing quark masses. They only enter
into the determination of the location of the minima at
$\chi_s(T)$ and $\bar\chi_0(T)$ via eq. (\ref{48AE}).
We indicate the vacuum properties for anomaly-dominated potentials
$(R_{an}=0.9)$ in table 2 for the same values of the input parameters
as used in table 1. Comparison between the
two tables provides an estimate of the effect of
nonzero degenerate  quark masses. We employ \cite{JW}
here a ``realistic'' average mass $m^2_{PS}=(390\ {\rm MeV})^2\approx
\frac{1}{3}(2M^2_K+M^2_\pi)$ which corresponds to the neglection
of $SU(3)$-violating mass splittings in the pseudoscalar
octet. We observe
a sizeable increase of $f=\sqrt{7/9}\bar\chi_0$ and a corresponding
increase in the baryon mass\footnote{Our choice
of the Yukawa coupling leads to too large
values of the baryon mass in vacuum for the potentials (B)(C).
This is only of little quantitative relevance
for the results presented in this note and can be corrected easily by
the choice of a smaller Yukawa coupling.}
$M_q=h_\chi(\bar\mu_\rho)\bar\chi_0$
and vector meson mass $\bar\mu_\rho=g(\bar\mu_\rho)\bar\chi_0$ in the
vacuum.

For all three potentials (A)(B)(C) we
find a first-order transition for  $m_{PS}$=390 MeV. The
critical temperature $T_c$ is increased substantially as compared
to the chiral limit, as can be inferred from table 2.

\begin{center}
\noindent {\bf Table 2:} First-order phase transition for
realistic average quark masses $(m_{PS}=390$ MeV).\\

\medskip
\begin{tabular}{|l|ccc|c|cc|c|}
\hline
\multicolumn{4}{|c|}{\rm vacuum}
&\multicolumn{4}{c|}{\rm critical temperature}
\\
\hline
$U_0$&$\bar\mu_\rho$&$f$&$M_\sigma$&$T_c$&$\bar\mu_\rho^{(SSB)}(T_c)$&
$M_q^{(SSB)}(T_c)$&$\tau(T_c)$\\
\hline
A&770&116&460&170&290&600&0.53\\
B&800&132&770&180&470&800&0.77\\
C&810&142&660&183&490&840&0.77\\
\hline
\end{tabular}
\end{center}

\medskip\noindent
The mass effect on the critical temperature
is moderate for our instanton-induced potential
(A) for which the increase of about 10 \% is roughly consistent
with a recent lattice simulation \cite{KLP}.
In this respect the instanton-induced potential computed in
\cite{instanton}
does apparently much better than the two polynomial
potentials (B)(C) that we use for comparison. We
point out, however, that the neglection of the quark mass effects
for the instanton-induced interaction leads to a substantial
uncertainty in the quark mass dependence at the present stage.

\section
{$Z_3$-Vortices}

In an attempt to find distinctive features of our scenario which could
be tested by lattice simultations, we may look at topologically stable
excitations. The
octet condensate in the low temperature phase indeed implies the
stability of macroscopic
$Z_3$-vortices. They are absent in the high temperature phase. In the
early universe, such
strings would have been produced as topological defects during the
QCD-phase transition.
The string
tension is typically of the order $\sigma \sim m^2_\rho$ such that $G
\sigma \approx
(m_\rho /M_p)^2$ is tiny. Observational consequences of the production
of typically
one string per horizon at the time of the phase transition are not
obvious - in particular
such strings are not candidates for seeds of a later galaxy formation.

As topological defects, the $Z_3$-vortices correspond to the nontrivial
homotopy group
$\pi_1(SU(3)/Z_3)=Z_3$. The color octet (and singlet) scalars transform
trivially under the
center of the color group, similar to the gluons. Any classical bosonic
field configuration
is therefore invariant under $Z_3$-transformations. Once the color group
gets "spontaneously
broken" by the octet condensate,
the $Z_3$-vortices become topologically stable. As an
example, consider a static vortex in the
z-direction with octet scalar field
\be \label{eq:field}
\chi_{ijab}=\frac{1}{2 \sqrt{6}}~\chi (r)~(\lambda_z)_{ab}~
\Big( v^{\dagger}(\varphi) \lambda_z v(\varphi) \Big) _{ji}
\ee
Here $v$ is given by a $\varphi$-dependent gauge transformation

\be
v (\varphi)=exp\Big(\frac{i}{\sqrt{3}}\varphi \lambda_8\Big)
\ee
such that a rotation around $2\pi$ corresponds to an element of $Z_3$

\be
exp(\frac{2\pi i}{\sqrt{3}}\lambda_8)=exp(\frac{2\pi i}{3})
\ee

The scalar field is therefore well defined and free of singularities
provided
$\chi (r)$ behaves properly at the origin, e. g. $\chi (r=0)=0$. For $r
\rightarrow \infty$
we assume that $\chi (r)$ approaches the expectation value $\chi_0$
which minimizes the
effective potential, $\chi (r \rightarrow \infty ) \rightarrow \chi_0$.
The potential
energy of the string is therefore concentrated in the core of the
string, typically of
radial size $m_{\rho}$. The nontrivial homotopy group tells us that no
smooth deformation can change the nontrivial behaviour for $r\rightarrow
\infty$.
In particular, no (singularity free) gauge transformation can "unwind"
the string.

Without the gauge fields, the gradient energy per length of the
configuration
(\ref{eq:field}), namely
$E_{grad}\sim \int drr^{-1} \partial_{\varphi} \chi^*_{ijab}
\partial_{\varphi}\chi_{ijab}
\sim \int drr^{-1}\chi^2(r)$,
would still diverge logarithmically for $r \rightarrow \infty$. We may
supplement a gauge field in
the $\varphi$-direction which becomes a pure gauge for
$r \rightarrow \infty \Big(a(r \rightarrow \infty )=1, a(r=0)=0\Big)$

\be
A_{\varphi}=-\frac{i}{g}\partial_{\varphi}
v^T~v^*a(r)=\frac{1}{\sqrt{3}g}~\lambda_8 a(r)
\ee
It is then easy to see that the covariant derivative vanishes for
$r \rightarrow \infty$
\be
D_{\varphi} \chi_{ijab}= \partial_{\varphi} \chi_{ijab}
-ig(A_{ik})_{\varphi} \chi_{kjab}+ig \chi_{ikab} (A_{kj})_{\varphi}
\rightarrow 0.
\ee
The gauge field strength vanishes as well in this limit.
We are therefore sure that the vortex has a finite string tension, the
precise value
depending on the shape of the functions $\chi (r)$ and $a(r)$ which have
to be determined by
solving the field equations with the appropriate boundary values for
$r=0$ and
$r \rightarrow \infty$.

Actually, the topological situation which leads to stable vortices gets
somewhat more
involved by the fact that both color and flavor groups are broken
simultaneously by the
octet condensate. A color rotation in the $Z_3$ element of $SU(3)_c$ can
be "unwound" by
a $\varphi$-dependent color-flavor-locked rotation. This shifts the
nontrivial topology from the
color sector to the flavor sector, more precisely to the nontrivial
center of the diagonal
flavor transformations from $SU(3)_L$ x $SU(3)_R$. The topological
stability due to the
nontrivial homotopy group $\pi_1$ is not affected. Since QCD is
invariant only under global
flavor transformations the vortex is, however, not invariant under
coordinate-dependent
color-flavor-locked transformations. The vanishing of the gauge
covariant kinetic term for
$r \rightarrow \infty$ occurs only if the  $\varphi$-dependent rotation
of the vacuum state is
associated to the color sector.

In this context it is useful to recall the global structure of the
symmetry group of QCD with three
massless quarks. For infinitesimal transformations the symmetry group is
\be
G=SU(3)_c \textup{ x }SU(3)_L \textup{ x } SU(3)_R \textup { x } U(1)_B
\ee
Global rotations in the center of $SU(3)_c$ and the center of the
diagonal vectorlike flavor
group $SU(3)_V$ correspond to identical phase rotations of the quark
fields. These
transformations belong also to the abelian group $U(1)_B$ which is
associated to conserved
baryon number. A direct consequence of this global group structure is
the triality rule:
Color representations in the class $(3, \overline{6}$, etc.) have baryon
number $B=1/3~ mod~
1$ and electric charge $Q=2/3~mod~1$ whereas the class (1, 8, etc) has
$B=1~mod~1$ and
$Q=1~mod~1$. This explains why the physical fermions are baryons with
$B=1$ and integer $Q$,
since all physical states are color singlets. (Some care is needed for
the correct
interpretation of the Higgs picture \cite{CW}.)

In a first look it may seem that the macroscopic $Z_3$-vortices are not
present in the
confinement picture of QCD, in contradiction to our assumption of
equivalence of the Higgs
and confinement pictures. This is, however, not obvious. In gluodynamics
(QCD without light
quarks) the $Z_N$-vortices are discussed as important configurations for
the understanding
of confinement \cite{vortex}. Lattice simulations could establish the
existence of macroscopic
vortices in the vacuum of three flavor QCD and their absence in the high
temperature phase.
Such a finding (or the contrary) may be interpreted as an important test
for our picture.

\section
{Conclusion}

In conclusion, we have presented here a rather simple picture of
the high-temperature phase transition in QCD. Both confinement and
chiral symmetry restoration are associated to the melting of a
color-octet
condensate at the critical temperature. The main phenomenological
features of our picture of the phase transition are the following:

(1) For three light quarks with equal mass a first-order phase
transition separates a low temperature ``hadronic phase''
from a high temperature ``quark-gluon phase''. In the chiral limit
of vanishing current quark masses the quarks and gluons are massless in
the high temperature phase where chiral symmetry is restored. Color
symmetry becomes a symmetry of the spectrum of
pseudoparticles above the critical temperature $T_c$. Below $T_c$
both chiral symmetry and color symmetry are spontaneously broken. Chiral
symmetry breaking generates a mass for the quarks which now appear
as baryons. Eight massless Goldstone bosons signal the global
symmetry breaking. Local color symmetry breaking gives a mass to
the gluons by the Higgs mechanism. The gluons are identified with
the octet of vector mesons $(\rho,K^*,\omega)$. The first-order
transition is therefore associated with a jump in the vector meson
mass (they become massless gluons for $T>T_c$) and the baryon
mass (they turn to massless quarks for $T>T_c$). Our picture
of gluon-meson duality and quark-baryon duality allows us
to describe the excitations relevant above and below $T_c$ by
the same fields. The first-order phase transition extends to nonzero
current quark mass, including values which lead in the vacuum
to  a ``realistic'' average mass for the $(\pi,K,\eta)$-pseudoscalars
$m_{PS}=390\ {\rm MeV}\approx\sqrt{(2M^2_K+M^2_\pi)/3}$.

(2) The effective multiquark interactions at low momentum
or the corresponding effective potential for $q\bar q$-bilinears
are presumably dominated by the axial anomaly arising from instanton
effects. We can then relate the critical temperature of a
strong first-order transition to the mass and decay constant of
the $\eta^\prime$-meson. For $m_{PS}=390$ MeV we find $T_c\approx$
170 MeV whereas the critical temperature is lower in the chiral
limit. For an anomaly dominated vacuum potential the prediction
for the chiral limit $(m_{PS}=0)$ is independent of many details
and found in the range $T_c\approx 130-160$ MeV.

(3) A dynamical picture for an instanton-induced color octet
condensate has been developed recently \cite{instanton}.
This has led to a computation of the octet potential which
involves the running gauge coupling as the only free parameter. The
characteristic features of the instanton-induced
vacuum effective potential
are depicted by the potential (A), eq. (\ref{2a}). We concentrate
in the following on  this estimate.
It is encouraging that the predictions of
the instanton-induced potential for
the critical temperature, namely $T_c$ = 170 MeV for $ m_{PS}
=$ 390 MeV and $T_c$ = 155 MeV for $m_{PS}=0$, agree well
with recent lattice simulations \cite{KLP}. One may interpret
this as a test for the idea of spontaneous color symmetry breaking
by instanton effects.

(4) For the equation of state in the quark-gluon phase we find at
$T_c$ a ratio between pressure and energy density $p/\epsilon\approx
0.13$ for $m_{PS}$ = 390 MeV and $p/\epsilon\approx0.19$ for
$m_{PS}=0$. This ratio approaches very rapidly the equation of state
of a relativistic  gas $(p/\epsilon=1/3)$ as $T$ increases beyond
$T_c$.

(5) In the hadronic phase the screening masses of baryons, vector
mesons and pseudoscalars show a strong temperature
dependence as $T$ approaches $T_c$ from below. Near $T_c$ the chiral
symmetry breaking contribution to
the average baryon mass $M_q\approx$ 600 MeV is only about one
half of the vacuum mass and approaches a typical constituent quark mass.
Also the Higgs contribution to the $\rho$-meson mass is
found substantially smaller than in the vacuum,
$\bar\mu_\rho(T_c)\approx$
300 MeV. For $T>T_c$ the effective chiral symmetry breaking
fermion mass jumps to a small value
$M_q\approx$ 40 MeV which is compatible for $m_{PS}=$ 390 MeV with
common estimates for the average current quark mass
$m_q=\frac{1}{3}(m_s+m_d+m_u)$. The average pseudoscalar mass
is found as $m_{PS}(T_c)\approx$ 180 MeV. It is therefore
substantially smaller than in the vacuum.

The present paper also contains a rather detailed discussion
of a mean field-type computation of the temperature dependence
of the effective potential. This has become necessary since a too
naive treatment of the scalar fluctuations has previously often led
to incorrect results near second-order or weak first-order
phase transitions. The reason is the complicated infrared
behavior of massless boson fluctuations in thermal equilibrium.
A second problem is the non-convexity of the perturbative
effective potential which leads naively to negative scalar mass terms.
Most previous work has simply left out the scalar fluctuations because
of the technical difficulties associated with these problems.
Nevertheless, the Goldstone boson fluctuations are important
for QCD at low temperature and near the phase transition, and
we therefore want to include them. Our prescriptions are based
on lessons from earlier more involved renormalization-group
investigations and should include the most dominant
effects. In particular, they account correctly for the low
temperature behavior as described by chiral perturbation theory
and are consistent with the physics of a second-order phase
transition if this occurs. Our treatment can easily be taken over for
other models.

Our approach to the high-temperature phase transition
in QCD still contains many uncertainties, which
have been discussed at various places in this paper. In particular, a
better understanding of the effective potential in the vacuum and
a separate treatment of the color octet and singlet condensates
would be most welcome and probably needed for a study of
finer aspects of the phase transition. This concerns, in particular,
the interesting question about the existence
and order of a phase transition in ``real QCD''
with different  strange quark and up/down quark
masses. An appropriate tool for this purpose seems to be
a renormalization group study similar to the succesful treatment
\cite{BJW} of the phase transition in the Nambu-Jona-Lasinio model.

Nevertheless, we find the simple overall picture
of the chiral and deconfinement transition rather convincing
and the quantitative results of a first rough computation
encouraging. We hope that this work can become a starting point
for a quantitative analytical understanding of the QCD
phase transition.

\section* {Appendix A: Running couplings}
\renewcommand{\theequation}{A.\arabic{equation}}
\setcounter{equation}{0}

In general, the effective masses $M$ appearing in the
$J$-integrals (\ref{8}) depend on the momentum $\vec q^2$. This could
be expressed through a momentum dependence of the effective gauge
and Yukawa couplings. We want to avoid here the complications of
momentum-dependent masses. Instead, the dominant effects of
the running couplings are taken into account by the choice of an
appropriate
renormalization scale $\mu$ in eq. (\ref{4}). Corrections to this
approximation are reflected in the term $\Delta J_G$ in eq. (\ref{7}).

For not too large $m$ the integrals (\ref{YY}) are dominated by
momenta $q^2\approx(\pi T)^2$. We therefore may
choose an appropriate temperature-dependent renormalization
scale
\be\label{11B}
\mu_T=\sqrt{\mu^2_\rho+\pi^2 T^2}\ ,\quad
\mu^2_\rho=g^2(\mu_\rho)\chi^2\ee
(Here $\mu^2_\rho=\mu^2_\rho(\mu=\mu_\rho)$ refers to the vacuum
with arbitrary $\chi$ corresponding to a possible presence of
``sources'' or quark mass terms.) This choice of $\mu_T$ is a valid
approximation for the fermion fluctuations and we therefore use
$h^2_\chi(\mu_T)$ in the argument of $J_F$ in eq. (\ref{7}).

For bosons the issue is more involved due to the infrared behavior
of classical statistics. The first two terms in a Taylor expansion,
$J_B(0)$ and $\partial J_B/\partial M^2_{|M^2=0}$, are dominated again
by momenta $\vec q^2\approx (\pi T)^2$. The remaining integral $\hat
J_B=J_B-J_B(0)-M^2
\frac{\partial J_B}{\partial M^2}_{|M^2=0}$ receives, however,
an essential contribution from the $n=0$ Matsubara frequency
which corresponds to classical statistics. The three-dimensional
momentum integrals of classcial statistics are more infrared-singular
than the four-dimensional integrals for the vacuum. In consequence,
the remaining integral $\hat J_B$ is dominated by momenta $\vec q^2
\approx M^2$. This leads in perturbation theory to the so-called
cubic term \cite{Dolan} which has played an important role in
the discussion of the electroweak phase transition \cite{CT}.
 Since $J_B$ is not analytic at $M^2=0$, a careful
treatment of the long-distance physics is mandatory if one aims
for quantitative precision. For first-order phase transitions
the momentum dependence of $M^2$ has only moderate effects for
the gauge boson fluctuations\footnote{For a second-order phase
transition
one may split the momentum integral in $J_B$ and use
$M^2=g^2_T(\mu_\rho)
\chi^2$ for $\vec q^2<(\mu_\rho/2)^2$. Here $g_T(\mu_\rho)$ obeys
an effective three-dimensional evolution for $\mu_\rho<\pi T$
\cite{ReuterW}.}
and we choose $M^2=g^2(\mu_T)\chi^2$ in $J_B$.

For the Goldstone boson fluctuations more care is needed if one
wants to be consistent with chiral perturbation theory for
low $T$ and the correct behavior at a second-order phase
transition. We first include
the gauge boson and fermion thermal fluctuations.
At this level the mass term for the Goldstone bosons
is given by
\be\label{23AA}
M^2_{G,0}=c_G\frac{\partial}{\partial(\chi^2)}(U_0
+\Delta_gU+\Delta_qU)\ee
In a second step we take the scalar fluctuations into account.
Then only the complete mass term $M^2_G$ (\ref{9}) vanishes
at the minimum of $U(\chi,T)$. The thermally corrected
(pseudo)particle mass term $M^2_G$ is relevant for momenta
$\vec q^2\stackrel{\scriptstyle<}{\sim}|M^2_G|$ and we
use $M^2_G$  for the argument of $J_G$.
On the other hand, the high momentum part of the integral $J_G$ is
dominated by momenta where the thermal mass corrections
from the scalar fluctuations are not yet
important. Therefore $M^2_{G,0}$ (\ref{23AA}) is relevant
in this momentum range. We
account for this by the correction
\be\label{20A3}
\Delta J_G=\frac{T^2}{24}(M^2_{G,0}-M^2_G)f_G(M_G^2/T^2)\ee
which effectively replaces $M^2_G$ by $M^2_{G,0}$ for the
term $\sim T^2$ which is dominated by momenta $\vec q^2\approx
(\pi T)^2$ if $T^2>|M^2_G|$. On the other hand, the
factor $f_G$ reflects the fact that all temperature
fluctuations and therefore also $\Delta J_G$ are exponentially
Boltzmann-suppressed for $T^2\ll M^2_G$. We choose here a simple
form ($\gamma=0.5$) which mimicks eq. (\ref{18AA}) for large $m^2$
\be\label{20.A4}
f_G(m^2)=\exp(\gamma-(\gamma^4+(m^2)^2)^{1/4})(1+\frac{(m^2)^2}{\gamma^4})
^{-1/8}\ee
These prescriptions may seem somewhat ad hoc and technical. They
find a deeper motivation from the study of the renormalization flow in
thermal field theories \cite{Renorm,BJW}. We will see below that they
guarantee
agreement with chiral perturbation theory for low temperature
and avoid unphysical singularities.

We finally have to specify the scale dependence of the gauge and
Yukawa coupling. For
the running gauge coupling we employ the perturbative $\beta$-function
in three-loop order in the $\overline{MS}$ scheme
\be\label{14}
\beta_g=\mu\frac{\partial}{\partial\mu}g=
-\beta_0\frac{g^3}{16\pi^2}-...\quad,\quad\beta_0=11-2N_f/3
\ee
Our normalization  $g$(770 MeV)=6 is suggested by the phenomenology
of $\rho$-decays into pions and $\mu^+\mu^-$ \cite{CW}.
For the Yukawa coupling we use the one-loop renormalization-group
equation
\be\label{16A}
\frac{dh_\chi^2}{d\ln\mu}=2\beta_h
h_\chi=\frac{a}{16\pi^2}h_\chi^4-\frac{b}{16\pi^2}
g^2h_\chi^2\ee
Treating also the gauge coupling
in one-loop order this has the solution
\be\label{16B}
h_\chi(\mu)=h_\chi(\mu_0)\left(\frac{g^2(\mu)}{g^2(\mu_0)}\right)
^{\frac{b}{4\beta_0}}\left(
1+\frac{ah^2_\chi(\mu_0)}{(2\beta_0-b)g^2(\mu_0)}\left[1-\left(
\frac{g^2(\mu)}{g^2(\mu_0)}\right)^{\frac{b}{2\beta_0}-1}\right]
\right)^{-1/2}
\ee
We use the approximation (\ref{16B}) with
$b=16$ such that we recover for
small $h_\chi$ the standard
anomalous dimension for the mass and take
somewhat arbitrarily\footnote{A more accurate treatment would keep the
color octet and singlet as separate fields. The evolution equations
for the two respective Yukawa couplings differ.} $a=1$. The
normalization
is fixed by eq. (\ref{4}), i.e.
$h_\chi$ (770 MeV)=1.15 GeV$/\chi_0$.

\section*{Appendix B: Second-order phase transition}
\renewcommand{\theequation}{B.\arabic{equation}}
\setcounter{equation}{0}

For an investigation of a possible
second-order phase transition in the chiral limit
we concentrate on the behavior
of $U$ near $\chi^2=0$
or small values of $\chi^2/T^2$. In the vicinity
of the critical temperature of a second-order (or weak
first-order) transition the Goldstone fluctuations play a
role in this region. The derivative of the
temperature-dependent effective potential becomes (in the range where
$U^\prime\geq0)$
\bear\label{15}
U^\prime(\chi,T)&=&U_0^\prime+\left(\hat g^2(\mu_T)+\frac{N_f}{4}\hat
h^2_\chi(\mu_T)+
\frac{N_f^2-1}{24}c_GU_0^{\prime \prime}\right)T^2\nonumber\\
&&-\left(\frac{3g^3\chi}{1-\beta_g/g}(\mu_\rho)+
\frac{N_f^2-1}{8}c_G^{3/2}U^{\prime \prime}\sqrt{U^\prime}\right)
\frac{T}{\pi}\ear
Here we have improved our treatment of the gauge boson
fluctuations by using $g(\mu_\rho)$ instead of $g(\mu_T)$ for
the low momentum fluctuations corresponding to classical
statistics (the term $\sim T$).
Let us denote by $T_0$ the temperature where
$U^\prime$ vanishes at $\chi=0$, i.e. $U^\prime(0,T_0)=0$. A
second-order phase
transition at $T_c=T_0$ occurs if $U^\prime(\chi,T_0)$ is strictly
positive for all $\chi>0$. In contrast, one has a first-order
transition if $U^\prime(\chi,T_0)$ is negative for small
$\chi>0$. Then the point $\chi=0$ corresponds to a maximum of
$U(\chi,T_0)$ and the critical temperature for the
first-order transition is below $T_0$.

Consider first the
potential (C) which is relevant
for the two-flavor case. Omitting for a
moment the Goldstone boson contributions and the running of the
gauge coupling $(\beta_g=0)$, the term
$\sim g^3\chi T$ in eq. (\ref{15}) would lead to a first-order
transition. This corresponds to the well-known ``cubic'' term in
the effective potential for the Higgs scalar in the electroweak
theory. As has been discussed extensively in connection with
the electroweak phase transition \cite{Sintra}, the infrared physics
of the gauge bosons responsible for this term has to be handled with
care.
In particular, the running of the gauge coupling is crucial for a
correct
picture. For a running $g$ the product $g\chi=\mu_\rho$ goes to
a constant $\Lambda(T)$ for $\chi\to 0$ and the nonanalyticity
in $\chi^2$ disappears. We note that $\chi$-independent
terms in eq. (\ref{15}) only influence the location of $T_0$. Denoting
$U^{\prime \prime}(0,T_0)=g_\chi$ one obtains for small $\chi$ and
$|T-T_0|$
an expansion, with constants
$c_1,c_2$ and $N_f^2-1=N$,
\be\label{16}
U^\prime=g_\chi\chi^2+c_1(T^2-T^2_0)+c_2(T-T_0)-\frac{Nc^{3/2}_G}{8\pi}
g_\chi T\sqrt{U^\prime}\ee
For $T\to T_0$ and $\chi\to0$ this yields \cite{RTW}
\be\label{17}
U^\prime(\rho,T)=\left[\frac{8\pi}{Nc_G^{3/2}T_0}(\chi^2+c_3(T-T_0))\right]^2
\ee
and describes \footnote{A first-order transition is possible if a new
minimum appears before the running $g$ has reached the behavior
$g\chi=const.$.} 
a second-order phase transition. The critical exponent
$\nu=1$ corresponds\footnote{This differs from
high-temperature perturbation theory or chiral
perturbation theory which would yield a ``mean field exponent''
$\nu=1/2$.} to the leading order of the $1/N$ expansion
in the effective three-dimensional theory \cite{RTW}. An improvement
of the description of the critical behavior (with more realistic
critical exponents) needs to incorporate the fluctuations of the
neglected scalar modes (see sect. 4) 
and can be done with the help
of modern renormalization group methods \cite{Renorm}. We conclude
that a second-order chiral phase transition for $N_f=2$ is compatible
with our picture, provided the infrared fluctuations of the gauge
bosons are treated in the appropriate way.

For three massless flavors the presence of an effective cubic term
$\sim\chi^3$ in the effective potential at zero temperature
(eq. (\ref{3})) changes the situation profoundly. This term 
reflects the axial anomaly and is induced
by instanton effects. The term $\sim-\chi$
always dominates the r.h.s.
of eq. (\ref{15}) for small enough values of $\chi$ since there
is no competing term linear in $\chi$. One infers the existence of
a first-order phase transition for three massless flavors.

The vicinity of the critical temperature of a second-order
or weak first-order transition is governed by universal behavior.
The universality class is characterized by the symmetry and the
representations which remain (almost) massless at
$T_c$. For two-flavor QCD it is not obvious which are the massless
scalar representations at a second-order phase transition. Furthermore,
the gauge bosons behave essentially as a free gas near the transition.
Even though their low momentum classical fluctuations may
acquire an effective mass, they will influence the nonuniversal
critical behavior. This influence of the gauge bosons is characteristic
for a simultaneous chiral and deconfinement transition and leads to
modifications of results obtained in quark-meson or Nambu-Jona-Lasinio
models \cite{BJW}. In view of the distance scales probed in lattice
simulations, we find it rather unlikely that critical behavior
in the standard $O(4)$-universality class will show up there. This 
is independent of the interesting question what would be the true 
universality calss of a second-order transition in two-flavor QCD,
as seen at very large correlation length close to $T_c$.

\section*{Acknowledgement}
The author thanks M. Alford and J. Berges for stimulating discussions.
This work was supported in part by the E.C. contract
ERB FMRX-CT97/0122 and by the Deutsche Forschungsgemeinschaft.


\end{document}